\documentclass[submission, Proceedings]{SciPost}

\hypersetup{
    colorlinks,
    linkcolor={red!50!black},
    citecolor={blue!50!black},
    urlcolor={blue!80!black}
}

\usepackage[bitstream-charter]{mathdesign}
\usepackage{subcaption}
\usepackage[capitalise]{cleveref}
\usepackage{siunitx}

\DeclareSymbolFont{usualmathcal}{OMS}{cmsy}{m}{n}
\DeclareSymbolFontAlphabet{\mathcal}{usualmathcal}

\newcommand{\msbar}{$\overline{\rm { MS}}$}
\newcommand{\mmsbar}{{\overline{\rm {MS}}}}
\DeclareMathOperator{\pos}{\rm POS}

\begin{document}

\begin{center}{\Large \textbf{
Can \msbar{} parton distributions be negative?
}}\end{center}

\begin{center}
Alessandro Candido\textsuperscript{1},
Stefano Forte\textsuperscript{1} and
Felix Hekhorn\textsuperscript{1$\star$}
\end{center}

\begin{center}
{\bf 1} Tif Lab, Dipartimento di Fisica, Universit\`a di Milano and\\ 
INFN, Sezione di Milano, Via Celoria 16, I-20133 Milano, Italy
\\
* felix.hekhorn@unimi.it
\end{center}

\begin{center}
\today
\end{center}

\definecolor{palegray}{gray}{0.95}
\begin{center}
\colorbox{palegray}{
  \begin{tabular}{rr}
  \begin{minipage}{0.1\textwidth}
    \includegraphics[width=22mm]{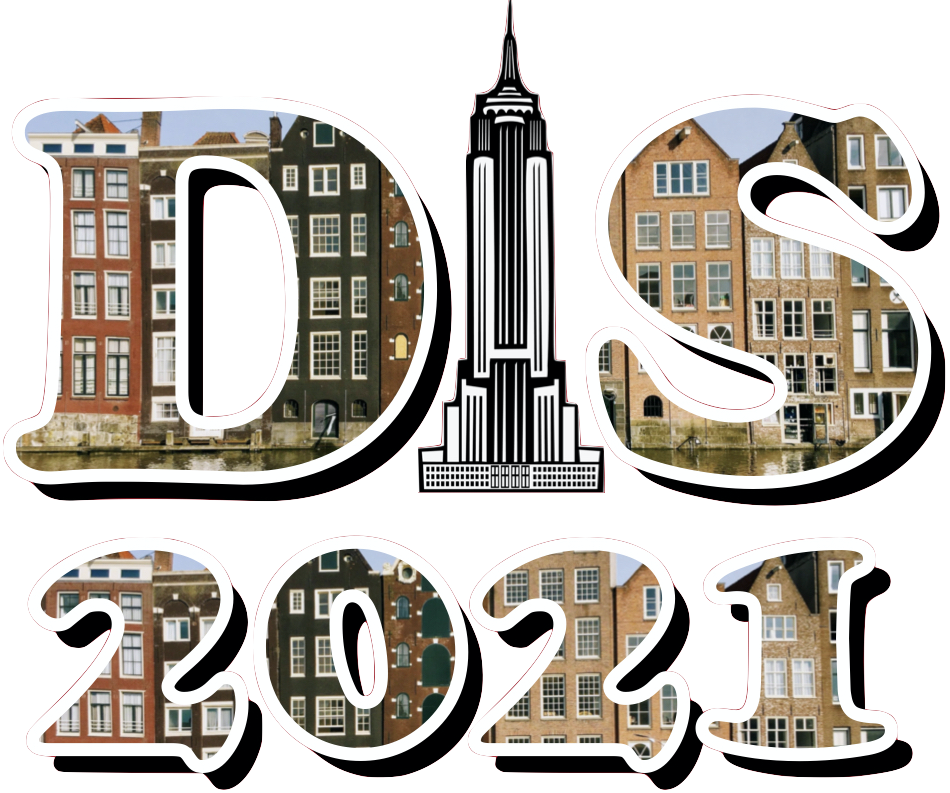}
  \end{minipage}
  &
  \begin{minipage}{0.75\textwidth}
    \begin{center}
    {\it Proceedings for the XXVIII International Workshop\\ on Deep-Inelastic Scattering and
Related Subjects,}\\
    {\it Stony Brook University, New York, USA, 12-16 April 2021} \\
    \doi{10.21468/SciPostPhysProc.?}\\
    \end{center}
  \end{minipage}
\end{tabular}
}
\end{center}

\section*{Abstract}
{\bf
    We review our recent paper, where we proved Parton distributions functions
    (PDFs) to be positive at NLO in the \msbar{} factorization scheme.
    We present some additional material that was useful in completing the steps
    for the actual proof, but we won't attempt to report on the full proof,
    but referring the reader to the paper itself.
}

\section{Introduction}
\label{sec:intro}
In our recent paper~\cite{Candido:2020yat} we tackled the long-standing problem of positive parton distribution functions (PDFs) head-on and finally gave a concise proof that, indeed, PDFs are perturbatively
positive at NLO. This has some immediate consequences for the actual fitting procedure as it affects the parametrizations of the PDFs.

The rest of the paper is structured as follows. We add some material about the
proof and a sketch of it (\cref{sec:proof}), and we show the impact on an
actual fit (\cref{sec:impact}).

\section{About the Proof}
\label{sec:proof}
In the following we quickly highlight the necessary step in order to prove the NLO statement.

First, we give a review of relevant coefficient functions of DIS and hadronic processes
(as prototypes of one and two-partons-involved processes) identifying the collinear substraction as the source of negativity.
Next, we present a possible physical scheme, i.e.\ a factorization scheme in
which the coefficient function for some candidate processes are reabsorbed at
all orders in the PDFs (like DIS scheme), in which PDFs are granted to be
positive because they coincide with actual observables.
Then, we construct a still positive scheme $\pos$, resembling \msbar{}, just
modifying the \msbar{} coefficient functions to make them positive as well.
We use the explicit positivity of coefficient functions, and exploit
perturbativity (and taking care of the large $x$ logarithms) to prove that at NLO
PDFs remain positive.
Finally, we compare the \msbar{} scheme to the constructed positive scheme $\pos$, in
order to prove the former to be positive as well.

\paragraph{Remark}
We would like to explicitly comment here that we are not interested in fitting
any of the intermediate schemes (e.g.\ $\pos$), they are just artifacts used
as intermediate steps to prove \msbar{} to be positive. 

\subsection{Positivity in $N$-space and $x$-space}
In \cite[Section 3.1]{Candido:2020yat} we have shown that the
positivity of the coefficient functions is a necessary and (perturbatively)
sufficient condition for PDF positivity.

During this process, we started working in Mellin space (or $N$-space) since convolutions are mapped onto ordinary products. The transformation also turns distributions into regular functions. We wanted to exploit the fact that we can control the sign of a multiplication and hence just use the inverse transformation to prove positivity.

We were mainly interested in the large $x$ region ($x \to 1$), since it is the
only region were \msbar{} coefficients are actually negative, and it is well known that
large-$N$ region asymptotically corresponds to the large $x$ region.

Nevertheless, we found remarkably that this is not enough for positivity: a
sufficiently positive function in $N$-space might map onto a function which is negative in the large $x$.

\paragraph{Explicit example}
An explicit example can be shown by making use of a simple polynomial in
$x$-space, and performing the Mellin transform analytically:

\begin{gather}
    f(x) = x (1-x) \left(x-\frac{7}{8}\right) \left(x-\frac{3}{4}\right)\\
    F(N) = \mathcal{M}[f(x)](N) = \frac{N^2-17 N+108}{32 (N+1) (N+2) (N+3) (N+4)}
\end{gather}

In this case the $N$-space function is positive in the range:
\begin{equation}
    \label{eq:posNregion}
    S_+ = (-\infty, -4) \cup (-3,-2) \cup (-1, \infty)
\end{equation}
and thus all $N > 1$. Nevertheless, the $x$-space function is not positive in the full interval $[0,1]$, but is only positive in:
\begin{equation}
    \left(0,\frac{3}{4}\right) \cup \left(\frac{7}{8}, 1\right)
\end{equation}

\begin{figure}[h]
\centering
\includegraphics[width=\textwidth]{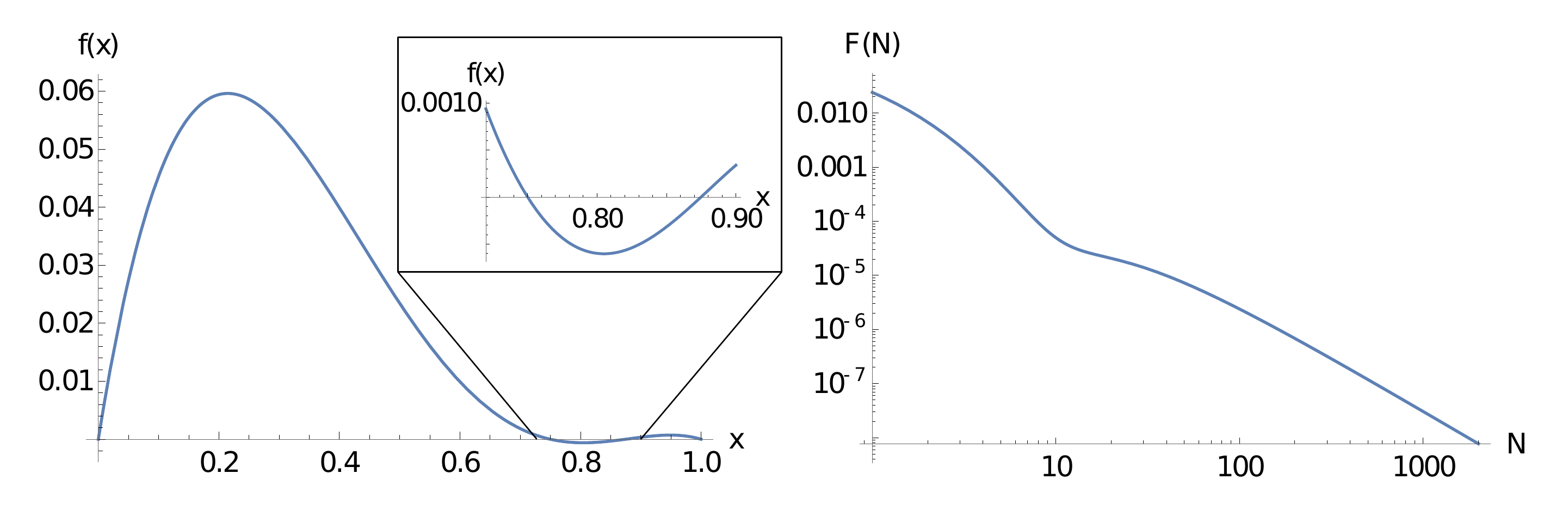}
\caption{Plotting our example function in $x$-space (left-hand side) and $N$-space (right-hand side)}
\label{fig:posN2posx}
\end{figure}

The former example is a specific case of the following functional form:
\begin{equation}
f(x)=x (1-x) (x-a) (x-b)
\end{equation}
that of course is negative in the interval $(a, b)$.

Moreover
\begin{equation}
    \forall b > 5-2 \sqrt{6} \sim 0.101 \quad \exists a : \forall N \in S_+(a,b) : F(N) > 0
\end{equation}
where $S_+(a,b)$ a set defined to $S_+$ \footnote{There is a lower bound
on $a$, in order to get that positive domain in $N$-space, such that it goes to
$1$ as $ b \to 1$. It is easy to compute but not relevant for the argument.}.
Than the function might be negative at arbitrarily large $x$ and still be positive for large $N$.

This required us to develop the whole argument directly in $x$-space, taking care
of distributions and convolutions directly.

\subsection{Considering the \msbar{} Scheme}
Following \cite[Section 3.2]{Candido:2020yat}, we consider the scheme change between the $\pos$ scheme and the actual \msbar{} scheme. In Mellin space we can denote the scheme change by a linear transformation:
\begin{equation}
f^{\mmsbar}(Q^2) = \left[\mathbb{I} + \frac{\alpha_s}{2\pi}K^{\pos} \right]^{-1} f^{\pos}(Q^2)
\end{equation}
where the PDFs $f(Q^2)$ are understood as a vector dimension over flavor space and the actual scheme transformation $K^{\pos}$ as a matrix in flavor space.

\begin{figure}[h]
\centering
\includegraphics[width=.5\linewidth]{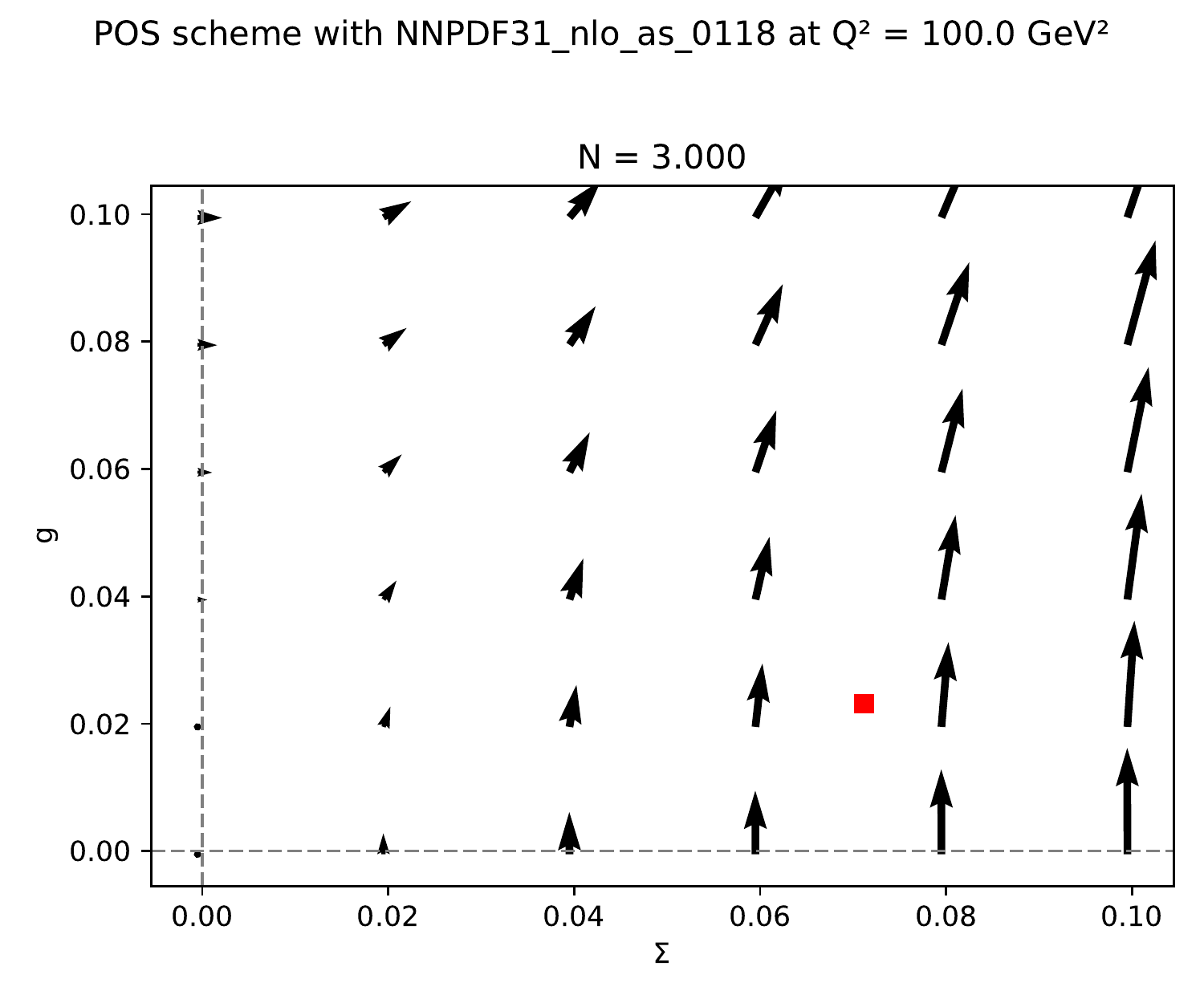}
\caption{Graphical representation of the scheme change between the $\pos$ and \msbar{} scheme: we plot the singlet moment against the gluon moment at $N=3$. The arrows indicate the shift of moments that the scheme change imposes. The red square denotes the reference value for the NNPDF3.1 PDF set\cite{NNPDF:2017mvq}}
\label{fig:quiver}
\end{figure}

While it was not clear to us from the beginning that the \msbar{} scheme actually \textit{is} a positive scheme, we were trying to construct a scheme change that explicitly enforces or breaks positivity. To that end we came up with the idea of a graphical representation for the scheme change as shown in \cref{fig:quiver}.

As an explicit example of such a plot, we plot here the third singlet moment against the third gluon moment and indicate by the arrows the effect that the scheme change has.
For reference we show the actual configuration that is realized by the central member of the NNPDF3.1 PDF set\cite{NNPDF:2017mvq} by the red square.
As can be seen, acutally all arrows are pointing into the first quadrant which means the scheme change from the $\pos$ to the \msbar{} scheme will make {\em{} any} configuration more positive.

\section{Impact onto NNPDF 4.0}
\label{sec:impact}
\begin{figure}[h]
\begin{subfigure}[b]{.5\linewidth}%
\centering\includegraphics[width=.9\linewidth]{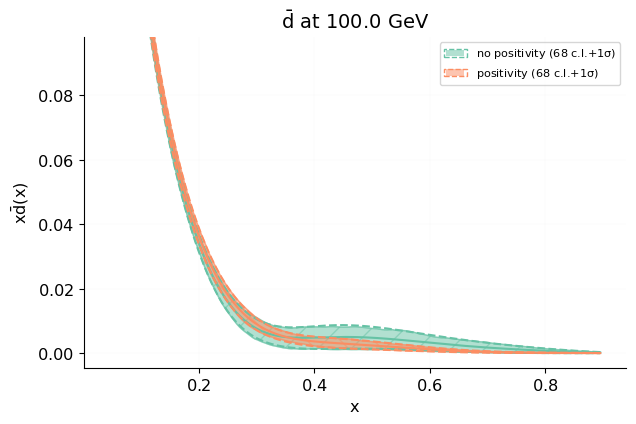}\caption{$\bar d(x)$ distribution}\end{subfigure}%
\begin{subfigure}[b]{.5\linewidth}%
\centering\includegraphics[width=.9\linewidth]{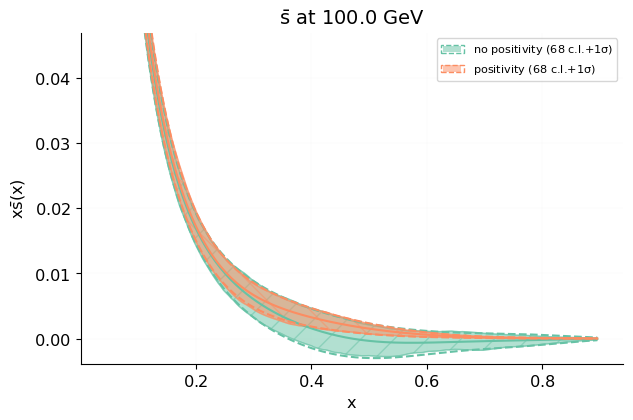}\caption{$\bar s(x)$ distribution}\end{subfigure}
\caption{Impact of imposing positivity on two selected PDF distributions at $\mu=\SI{100}{\GeV}$}
\label{fig:fit}
\end{figure}

The strict positivity of PDFs in the flavor basis is implemented in the forthcoming NNPDF4.0 release~\cite{nnpdf40} by the use of Lagrange mulipliers. In \cref{fig:fit} we demonstrate the impact of imposing this constraint by two exemplary PDFs. Indeed, as expected, PDFs can and will become negative in the large $x$ region and with the newly imposed constraint we avoid this unphysical region. Furthermore, we even see a reduced error band when imposing positivity.

\section{Conclusion}
We observe that posivitiy in momentum fraction space space ($x$-space) and Mellin moment space ($N$-space) do \textit{not} coincide. Furthermore, we give a graphical representation of the scheme change that is used in the actual proof and which demonstrates the actual positivity in a more visual sense.

The positivity of \msbar{} PDFs leads to an improved fitting methodology due to the reduced function space that has to be explored. This has a sizable impact on the fitting procedure, but it remains unclear for the moment whether there exists a yet stricter condition that can be imposed to enforce positive, physical cross sections.

However, we want to point out again that positivity is neither a necessary nor a sufficient condition for obtaining physical PDFs: it is not necessary because the convolution a partially negative PDF with a sufficiently good coefficient functions might still yield positive, physical cross sections. Yet it is not a sufficient condition because due to numerical or methodological inefficientcies (e.g.\ the use of perturbation theory itself) also the convolution of strictly positive PDFs with sufficiently bad coefficient functions might yield negative, unphysical cross sections.

\section*{Acknowledgements}
We acknowledge the members of the NNPDF and N3PDF collaborations for careful reading
and discussions.

\paragraph{Funding information}
This work is  supported by the European Research Council under
the European Union's Horizon 2020 research and innovation Programme
(grant agreement n.740006).

\bibliography{refs.bib}

\nolinenumbers

\end{document}